\documentclass[12pt,twoside]{article}
\usepackage{amssymb}
\usepackage{amsmath}
\usepackage[dvips]{graphicx}
\begin{document}
\begin{center}
{\large Quantum Potential Via General Hamilton - Jacobi Equation}

\vspace{0.6cm} {\bf Maedeh. Mollai}$^{1}$~,{\bf Mohammad. Razavi}$^{1}$,~{\bf Safa. Jami}$^{1}$
and  {\bf Ali Ahanj}$^{2,3}$\footnote{Electronic address: ahanj@ipm.ir; ~~Fax: 00985116635999} ,\\
\vspace{0.2cm} {\it ${}^1$ Department of Physics, Islamic Azad University, Mashhad Branch, Mashhad , Iran.}\\
\vspace{0.2cm} {\it ${}^2$ Department of Physics, Khayyam Higher Education Institute (KHEI), Mashhad, Iran. }\\
\vspace {0.2cm} {\it ${}^3$ School of Physics, Institute for Research in Fundamental sciences (IPM), Tehran, Iran.}

\end{center}

\begin{abstract}
 In this paper, we sketch and emphasize  the automatic emergence of a quantum potential (QP) in general Hamilton-Jacobi equation via commuting relations, quantum canonical transformations and without the straight effect of wave function. The interpretation of QP in terms of independent entity is discussed along with the introduction of quantum kinetic energy. The method has been extended to relativistic regime, and same results have been concluded.
\end{abstract}
\vspace{0.3cm}
\begin{flushright}
\end{flushright}
\begin{flushleft}
\end{flushleft}

\section{Introduction }
The de Broglie-Bohm's pilot wave theory, introducing the concept of quantum potential (QP), provided a causal, and at the same time nonlocal, description for quantum phenomena\cite{holland}. According to the theory, the particle is assumed to be in a ``thermal" bath, provided by a background of vacuum fluctuations \cite{bohm},  which affects on particles via QP \cite{holland}.\\ QP is a direct consequence of simply rewriting the Scher$\ddot{o}$dinger equation in its real and imaginary parts under polar decomposition of the wave function\cite{elitzur}. It is the term that distinguishes classical mechanics from quantum mechanics; but it has provoked some hostile criticism to the Bohm approach  and some physicists feel the QP is somehow artificial and should be avoided \cite{elitzur}. In some sense they are implicitly following on Heisenberg  who argued that the approach needed ``some strange QPs introduced in ad hoc manner by Bohm" \cite{hiley}. In other word, the ad hoc but necessary appropriation of the Schr$\ddot{o}$dinger equation as the equation for the Bohmian wave has an unattractively opportunist air to it \cite{polking}.\\ In this work, we come to the theory's defense by providing an innovative approach to QP via general Hamilton-Jacobi (HJ) equation, canonical transformation and without straight effect of wave function. Hamilton-Jacobi formalism is elegant, general and powerful; it has a direct connection with quantum theory, and is conceptually clear. It is well known that the formal relationship between quantum mechanics and classical mechanics is expressed in the analogy between commutator brackets and Poisson brackets, and between Heisenberg's equations of motion and Hamilton's equations of motion \cite{jordan}. Canonical transformations play a central role in the classical mechanics \cite{goldstein}. The presence of noncommuting operators makes quantum canonical transformations different from the classical ones \cite{jordan}. Owing to the formal similarities between classical and quantum mechanics, quantum canonical transformations closely resemble their classical counterparts \cite{roncadelli}.\\The ultimate outcome includes QP and a new term that we call quantum kinetic energy (QK). We try to show that QP is not ad hoc and merely reflects  the wave function \cite{maudlin} and along with  QK they are located at the heart of HJ and they are produced due to the non-commuting operators in the HJ. Also  we have extended our method to relativistic regime, and the same results have been obtained.\\ Traditionally, time enters quantum mechanics as a  parameter rather than a dynamical operator. In this way quantum mechanics differs from special relativity where time and space coordinates are treated on an equal footing \cite{provonic}.The concept of the ``time operator" is strongly connected with the time-energy uncertainty relation \cite{aharanov}. On the other hand, the criticism posed by Pauli \cite{pauli} , although it is not rigorous, that the time operator can not necessarily be defined for all quantum systems without contradiction. Furthermore the physical meaning of the time operator, if any, still remains unclear(see, e.g., \cite{provonic, wang, miyamoto, razavi,nikolic} for reviews). In this paper, since we have extended the method to the relativistic regime, to preserve the symmetry, we consider time as an operator: $ \hat{t}|t>=t|t>$.\\ The paper is organized as follows. In Sec. II, we  obtain the general Hamilton-Jacobi equation in nonrelativistic regime and will introduce quantum potential and quantum kinetic energy. Then, we will extend our approach to relativistic regime in Sec III. Finally,Sec IV reports  the conclusion.
\section{Non-Relativistic Quantum Potential}
In the Hamiltonian formulation, the momenta are independent variables on the same level as the generalized coordinates. The concept of transformation of coordinates must therefore be widened to include the simultaneous transformation of the independent coordinates and momenta $\hat{q}_i$ ,$\hat{p}_i$ to a new set $\hat{Q}_i$ ,$\hat{P}_i$ with (invertible) equations of transformation:
 \begin{eqnarray}
 \label{qp}
\hat{Q}_i=\hat{Q}_i(\hat{q},\hat{p},\hat{t})\nonumber\\
\hat{P}_i=\hat{P}_i(\hat{q},\hat{p},\hat{t}).
\end{eqnarray}
 In developing Hamiltonian mechanics, only those transformations can be of interest for which the new $\hat{Q}_i$ ,$\hat{P}_i$   are canonical coordinates. This requirement will be satisfied provided there exist some function $\hat{K}(\hat{Q},\hat{P},\hat{t})$, so called new-Hamiltonian, such that $\hat{ K}$ is related to the old-Hamiltonian and to the generating function by the equation:
 \begin{equation}
 \label{k=h}
\hat{K}(\hat{Q},\hat{P},\hat{t})=\hat{H}(\hat{Q},\hat{P},\hat{t})+\frac{\partial\hat{F}(\hat{Q},\hat{P},\hat{t})}{\partial \hat{t}}.
\end{equation}
The generating function producing the desired transformation is Hamilton's principal function, $\hat{F}(\hat{Q},\hat{P},\hat{t})$ and $\frac{\partial\hat{F}(\hat{Q},\hat{P},\hat{t})}{\partial t}$  provides the connection between the new-Hamiltonian, $\hat{K}$, and the old one,  $ \hat{H}$  \cite{goldstein}.  Note that, here, we choose the set $(\hat{q},\hat{Q})$ , from four possible sets of independent canonical variables \cite{explain}.
When the Hamiltonian is any general function of $\hat{H}(\hat{Q},\hat{P},\hat{t})$, we seek canonical transformations to new variables such that all the coordinates and momenta  $\hat{Q}_i$,  $\hat{P}_i$ are constants of motion. To meet these requirements, it is sufficient to demand that the new-Hamiltonian shall vanish identically, $ \hat{K}=0$. Under these conditions, the generating function producing the desired transformation is Hamilton's principal function, $\hat{S}(\hat{Q},\hat{P},\hat{t})$, satisfying the Quantum Hamilton-Jacobi (QHJ) partial differential equation \cite{goldstein}:
 \begin{equation}
 \label{h+s}
\hat{H}(\hat{Q},\hat{P},\hat{t})+\frac{\partial\hat{S}(\hat{Q},\hat{P},\hat{t})}{\partial \hat{t}}=0.
\end{equation}
As emphasized by Jordan and Dirac, the resulting operator order ambiguity should be fixed by enforcing well-ordering: operators represented by capital letters should always stay to the right of those labeled by lower case letters \cite{elitzur}. This means that $\hat{S}(\hat{Q},\hat{P},\hat{t})$  should have the structure of:
\begin{equation}
\label{s}
\hat{S}(\hat{Q},\hat{P},\hat{t})=\Sigma_{\alpha} f_{\alpha}(\hat{q},\hat{t}) g_{\alpha}(\hat{Q},\hat{t})
\end{equation}
where $f_{\alpha}(\hat{q},\hat{t})$ and $ g_{\alpha}(\hat{Q},\hat{t})$   are suitable functions. Other canonical transformations can be written as ($1\leq i \leq N$):
 \begin{eqnarray}
 \label{p}
\hat{p}_i = \frac{\partial\hat{S}(\hat{Q},\hat{P},\hat{t})}{\partial \hat{q}_i} \\
\hat{P}_i = \frac{\partial\hat{S}(\hat{Q},\hat{P},\hat{t})}{\partial \hat{Q}_i}.
\end{eqnarray}
So QHJ could be written as:
 \begin{equation}\label{hs2}
\hat{H}(\hat{q},\frac{\partial\hat{S}(\hat{q},\hat{Q},\hat{t})}{\partial \hat{q}_i},\hat{t})+\frac{\partial\hat{S}(\hat{q},\hat{Q},\hat{t})}{\partial \hat{t}}=0
\end{equation}
as in classical mechanics, the operator Hamilton-Jacobi equation, Eq. ($\ref{hs2}$), provides an independent formulation of the theory \cite{roncadelli}.
Our aim is to show that the quantum potential (QP) is located at the heart of QHJ, and it is produced only due to the non-commuting operators in the generating function. We apply general Weyl-ordering throughout QHJ equation. General Weyl-ordered Hamiltonian is \cite{roncadelli, lee}:
 \begin{eqnarray}\label{weyl}
 H(\hat{q},\hat{p},\hat{t})&=&\frac{1}{4}a_{ij}(\hat{q},\hat{t}) \hat{p}_i\hat{p}_j +\frac{1}{2}\hat{p}_i a_{ij}(\hat{q},\hat{t})\hat{p}_j +\frac{1}{4} \hat{p}_i\hat{p}_ja_{ij}(\hat{q},\hat{t})\nonumber\\&+&\frac{1}{2}b_{i}(\hat{q},\hat{t}) \hat{p}_i+
 \frac{1}{2}\hat{p}_i b_{i}(\hat{q},\hat{t})+c(\hat{q},\hat{t}),
\end{eqnarray}
where  $a_{ij}(\hat{q},\hat{t}) $ , $ b_{i}(\hat{q},\hat{t})$  and  $c(\hat{q},\hat{t})$ are functions of $\hat{q}_k$  and  $\hat{t}$ and summation over repeated Latin indices is understood. Note that, as  $\hat{p}_i\hat{p}_j $  and  $\frac{\partial\hat{S}}{\partial \hat{q}_i}$ are comparably equivalent, we should put a coefficient for the latter and since here $\hat{t}$ is considered as an operator,  $\frac{\partial\hat{S}}{\partial \hat{t}}$ should be Weyl-ordered. Accordingly, by employing general Weyl-ordering and the shorthand $\hat{S}\equiv \hat{S}(\hat{q},\hat{Q},\hat{t})$ , Eq. (\ref{hs2}) reads:
 \begin{eqnarray}
 \label{asli1}
 &&\frac{1}{4}a_{ij}(\hat{q},\hat{t}) \frac{\partial \hat{S}}{\partial \hat{q}_i}\frac{\partial \hat{S}}{\partial \hat{q}_j}+ \frac{1}{2}\frac{\partial \hat{S}}{\partial \hat{q}_i}a_{ij}(\hat{q},\hat{t})\frac{\partial \hat{S}}{\partial \hat{q}_j}+ \frac{1}{4} \frac{\partial \hat{S}}{\partial \hat{q}_i}\frac{\partial \hat{S}}{\partial \hat{q}_j} a_{ij}(\hat{q},\hat{t})\nonumber\\&&+ \frac{1}{2}b_{i}(\hat{q},\hat{t}) \frac{\partial \hat{S}}{\partial \hat{q}_i}+
 \frac{1}{2}\frac{\partial \hat{S}}{\partial \hat{q}_i} b_{i}(\hat{q},\hat{t})+\frac{a(\hat{q},\hat{t})}{2}\frac{\partial \hat{S}}{\partial \hat{t}}+
 \frac{\partial \hat{S}}{\partial \hat{t}}\frac{a(\hat{q},\hat{t})}{2}\nonumber\\&&+c(\hat{q},\hat{t})=0.
\end{eqnarray}
By replacing $ a_{ij}(\hat{q},\hat{t})=\frac{a(\hat{q},\hat{t})}{2m}A_{ij}$  and sandwiching between $<q,t|$ and $|Q,t>$, we have:
\begin{eqnarray}
\label{asli2}
&&\frac{a(q,t)}{4m}A_{ij}<q,t| \frac{\partial \hat{S}}{\partial \hat{q}_i}\frac{\partial \hat{S}}{\partial \hat{q}_j}|Q,t>+ \frac{1}{2m}<q,t|\frac{\partial \hat{S}}{\partial \hat{q}_i}a(\hat{q},\hat{t})A_{ij}\frac{\partial \hat{S}}{\partial \hat{q}_j}|Q,t>\nonumber\\ &&+\frac{1}{4m} <q,t|\frac{\partial \hat{S}}{\partial \hat{q}_i}\frac{\partial \hat{S}}{\partial \hat{q}_j} a(\hat{q},\hat{t})A_{ij}|Q,t>+ \frac{1}{2}b_{i}(q,t)
 <q,t|\frac{\partial \hat{S}}{\partial \hat{q}_i}|Q,t>\nonumber\\&&+\frac{1}{2}<q,t| \frac{\partial \hat{S}}{\partial \hat{q}_i}b_i(\hat{q},\hat{t})
|Q,t>+\frac{1}{2}a(q,t)<q,t|\frac{\partial \hat{S}}{\partial \hat{t}}|Q,t>+\nonumber\\&&  \frac{1}{2}<q,t| \frac{\partial \hat{S}}{\partial \hat{t}}a(\hat{q},\hat{t})|Q,t>+ c(q,t)<q,t|Q,t>=0.
\end{eqnarray}
By using the commutator's algebra:
\begin{eqnarray}
\frac{\partial \hat{S}}{\partial \hat{q}_i}G(\hat{q},\hat{t})&=& G(\hat{q},\hat{t})\frac{\partial \hat{S}}{\partial \hat{q}_i}-i\hbar\frac{\partial \hat{G}}{\partial \hat{q}_i}\\\frac{\partial \hat{S}}{\partial \hat{t}}G(\hat{q},\hat{t})&=& G(\hat{q},\hat{t})\frac{\partial \hat{S}}{\partial \hat{t}}-i\hbar\frac{\partial \hat{G}}{\partial \hat{t}},
\end{eqnarray}
and by taking $G(\hat{q},\hat{t}) \equiv a(\hat{q},\hat{t}) $, $G(\hat{q},\hat{t}) \equiv b_i(\hat{q},\hat{t})$ and $G(\hat{q},\hat{t}) \equiv \frac{\partial S(\hat{q},\hat{t})}{\partial \hat{q}_j}$, we can rewrite Eq($\ref{asli2}$)as:
\begin{eqnarray}
\label{asli3}
&&\frac{a(q,t)}{m}A_{ij} <q,t|\frac{\partial \hat{S}}{\partial \hat{q}_i}
\frac{\partial \hat{S}}{\partial \hat{q}_j}|Q,t>+2a(q,t)<q,t|\frac{\partial \hat{S}}{\partial \hat{t}}|Q,t>+\nonumber\\&&2\left(b_i(q,t)-\frac{i\hbar}{2m}\frac{\partial a(q,t)}{\partial q_j}A_{ij}\right)<q,t|\frac{\partial \hat{S}}{\partial \hat{q_i}}|Q,t>+ \nonumber\\&& \left(2c(q,t)-i\hbar \frac{\partial \hat{b_i(q,t)}}{\partial \hat{q_i}} - \frac{\hbar^2}{4m}\frac{\partial ^2a(q,t)}{\partial q_i \partial q_j}A_{ij}-i \hbar \frac{\partial a(q,t)}{\partial t}\right)\nonumber\\&& <q,t|Q,t>=0.
\end{eqnarray}

At this point, we denote by $S \equiv S(q,Q,t)$   the c-number function that uniquely produces$S(\hat{q},\hat{Q},\hat{t})$   by enforcing well-ordering and the substitution $q \longrightarrow\hat{q}$, $Q \longrightarrow\hat{Q}$  \cite{roncadelli} and $t \longrightarrow\hat{t}$ . Explicit use of equation ($\ref{s}$) yields:
\begin{eqnarray}
\label{11}
<q,t|\hat{S}|Q,t>&=&S<q,t|Q,t>\\
<q,t|\frac{\partial\hat{S}}{\partial\hat{t}}|Q,t>&=&\frac{\partial S}{\partial t}<q,t|Q,t>\\
<q,t|\frac{\partial\hat{S}}{\partial\hat{q}_i}|Q,t>&=&\frac{\partial S}{\partial q_i}<q,t|Q,t>
\end{eqnarray}
and furthermore
\begin{eqnarray}
\label{12}
<q,t|\frac{\partial \hat{S}}{\partial \hat{q}_i}\frac{\partial \hat{S}}{\partial \hat{q}_j}|Q,t>=
( \frac{\partial S}{\partial q_i}\frac{\partial S}{\partial q_j} -i\hbar\frac{\partial ^2 S}{\partial q_i \partial q_j})<q,t|Q,t>.\nonumber\\
\end{eqnarray}
By substituting these in Eq ({$\ref{asli3}$}), we obtain:
\begin{eqnarray}
\label{asli4}
&&\frac{a(q,t)}{m}A_{ij} \left(\frac{\partial S}{\partial q_i}
\frac{\partial S}{\partial q_j}-i\hbar  \frac{\partial ^2 S}{\partial q_i \partial q_j}\right)+ 2 a (q,t) \frac{\partial S}{\partial t}+\nonumber\\ && 2 \left(b_i(q,t)-\frac{i\hbar}{2m}\frac{\partial a(q,t)}{\partial q_j}A_{ij}\right) \frac{\partial S}{\partial q_i}+\nonumber\\&& \left(2c(q,t)-i\hbar \frac{\partial b_i(q,t)}{\partial q_i} - \frac{\hbar^2}{4m}\frac{\partial ^2a(q,t)}{\partial q_i \partial q_j}A_{ij} -i \hbar \frac{\partial a(q,t)}{\partial t}\right) \nonumber\\&&=0.
\end{eqnarray}
This derivation makes it natural to regard Eq.($\ref{asli4}$) as the c-number QHJ equation associated with the operator QHJE described by the quantum Hamiltonian.
The physical significance of Eq.($\ref{asli4}$) becomes clear by separating imaginary and real parts :
\begin{eqnarray}
\label{real}
&&\frac{A_{ij}(q,t)}{2m}\frac{\partial S}{\partial q_i}\frac{\partial S}{\partial q_j}-\frac{\hbar ^2}{8m}\frac{1}{a(q,t)}\frac{\partial ^2 a(q,t)}{\partial q_i \partial q_j}A_{ij}(q,t) \nonumber\\&&  +\frac{b_i (q,t)}{a(q,t)}\frac{\partial S}{\partial q_i}+\frac{c(q,t)}{a(q,t)} +\frac{\partial S}{\partial t}=0
\end{eqnarray}
\begin{eqnarray}\label{imagin}
 \frac{\partial}{\partial q_j}\left( \frac{a(q,t)A_{ij}(q,t)}{m}\frac{\partial S}{\partial q_i}\right)+\frac{\partial a(q,t)}{\partial t} + \frac{\partial b_i (q,t)}{\partial q_i}=0.\nonumber\\
\end{eqnarray}
Equations ($\ref{real}$) and ($\ref{imagin}$) are a pair of coupled partial differential equations. In the special case which $A_{ij}$ is an unit diagonal tensor and by considering $i=j$ elements, eqs ($\ref{real}$) and ($\ref{imagin}$) read:
\begin{eqnarray}
\label{akhar2}
\frac{\partial S}{\partial t}+\frac{(\nabla S)^2}{2m}-\frac{\hbar ^2}{8m}
\frac{\nabla ^2 a(q,t)}{a(q,t)}+\frac{b(q,t).\nabla S}{a(q,t)}+\frac{c(q,t)}{a(q,t)}=0\nonumber\\
\end{eqnarray}
\begin{equation}
\label{akhar1}
\frac{\partial a (q,t)}{\partial t}+ \nabla . \left( a(q,t) \frac{\nabla S(q,Q,t)}{m}\right) +\nabla . b(q,t)=0.
\end{equation}
Now, recall that the two equations defining the Bohm approach emerge from the Schrodinger equation by simply writing the wave function in polar form:
$\psi=R exp(iS/\hbar)$. Then the resulting equation is split into its real and imaginary parts and we find:
  \begin{eqnarray}
  \label{akhar3}
\frac{\partial R^2 (q,t)}{\partial t}+ \nabla  \left(R^2(q,t) \frac{\nabla  S}{m}\right) =0
\end{eqnarray}
which gives us a conservation of probability equation. The real part of the Schr$\ddot{o}$dinger equation gives:
\begin{equation}
\label{akhar4}
\frac{\partial S }{\partial t}+\frac{(\nabla S)^2}{2m}-\frac{\hbar ^2}{2m}\frac{\nabla^2 R(q,t)}{R(q,t)}+V=0.
\end{equation}
This equation  resembles the Hamilton-Jacobi equation except that it contains an extra term
$QP=-\hbar ^2/2m(\nabla^2 R/R)$ , which has been called the quantum potential since it is this term that distinguishes classical mechanics from quantum mechanics. If we identify  $\nabla S$ with momenta and regard $QP$ as a new quality of energy only playing a role in quantum process, then we can regard Eq($\ref{akhar2}$) as an expression of conservation of energy \cite{elitzur}.
Now,  the comparison of Eqs. ($\ref{akhar1}$) and ($\ref{akhar2}$) with ($\ref{akhar3}$) and ($\ref{akhar4}$), makes it natural to regard $a(q,t)=R^2(q,t)$  and $\nabla . b(q,t)=0$ ,
 ( where $ \vec{b}(q,t)=R^2(q,t)\vec{V}(q,t) $   is a vector potential) and  $ c(q,t)=R^2(q,t) V(q,t)$; thereby we obtain:
\begin{eqnarray}
\label{last1}
&&\frac{\partial S }{\partial t}+\frac{(\nabla S)^2}{2m}-\frac{\hbar ^2}{4m}\frac{\nabla^2 R(q,t)}{R(q,t)}-\nonumber\\&&\frac{\hbar ^2}{4m}\left(\frac{\nabla R (q,t)}{R(q,t)}\right)^2+ \vec{V}(q,t).\nabla S(q,Q,t)+V(q,t)=0\nonumber\\
\end{eqnarray}
\begin{eqnarray}
\label{last2}
\frac{\partial R^2 (q,t)}{\partial t}+ \nabla. \left(R^2(q,t) \frac{\nabla  S}{m}\right) =0.
\end{eqnarray}
We stress that this result holds true even for solutions  $S(q,Q,t)$ of the operator QHJ equation that are independent of $Q$ , since all equations from ($\ref{asli1}$) onward could have been multiplied by $\int dQ \phi (Q)$  , with $\phi(Q)$  being arbitrary \cite{roncadelli}.
Since QP is defined as$QP=-\hbar ^2/2m(\nabla^2 R/R)$, we name $QK=-\hbar ^2/2m\left(\nabla R/R\right)^2 $ as quantum kinetic energy (QK) which is an external energy of particle due to ``vacuum fluctuations" \cite{grossing}. So we can rewrite Eq.($\ref{last1}$) as
\begin{eqnarray}
\label{last3}
&&\frac{\partial S }{\partial t}+\frac{(\nabla S)^2}{2m}+\frac{1}{2}(QP+QK)\nonumber\\&&+ \vec{V}(q,t).\nabla S+V(q,t)=0
\end{eqnarray}
where we could read $QP+QK=-\frac{\hbar ^2}{2m}\frac{\nabla.(R\nabla R)}{R^2}$ .
Eq. ($\ref{last3}$) is general QHJ equation which includes Schr$\ddot{o}$dinger equation. The approach leads to the spontaneous appearance of ``quantum potensial" (QP) and ``quantum kinetic energy" (QK) via a mere quantum approach, without the straight effect of wave function. So we have proved that QP is not ad hoc and is merely a reflection of the wave function\cite{maudlin}. But along with QK they  are produced due to the noncommuting operators in the HJ equation.
\section{Relativistic Quantum Potential}
The above method could be extended to the relativistic regime, by starting from relativistic Hamilton-Jacobi equation (RHJ):
 \begin{equation} H(\hat{q}^{\mu},\hat{p}^{\mu})=0.
 \end{equation}
The relativistic formulation of HJ theory is simpler than the conventional non relativistic version, indicating that the relativistic formulation unveils a natural and general structure of mechanical system \cite{rovelli}. By applying Weyl-ordering to the general Hamiltonian, we obtain:
\begin{eqnarray}
\label{r1}
&&c(\hat{q}^{\mu}) + \frac{1}{2} \left( b_{\mu}(\hat{q}^{\mu})\hat{p}^{\mu}+\hat{p}^{\mu}b_{\mu}(\hat{q}^{\mu})\right) +\frac{1}{4}
 a(\hat{q}^{\mu})\hat{p}^{\mu}\hat{p}_{\mu} \nonumber\\&&+\frac{1}{2}\hat{p}^{\mu}a(\hat{q}^{\mu})\hat{p}_{\mu}+\frac{1}{4}\hat{p}^{\mu}\hat{p}_{\mu}a(\hat{q}^{\mu})=0
 \end{eqnarray}
 By using $ \hat{p}^{\mu} = - \frac{\partial S ( \hat{q},\hat{Q},\hat{t})}{\partial \hat {q}^{\mu}} $ and the canonical communication relation $[ G(\hat{q}^\mu), \partial^\mu \hat {S}] = i \hbar \partial_\mu G(\hat{q}^\mu) $ , and sandwiching Eq.($\ref{r1}$)between $<q|$ and $|Q>$ and doing the same calculations like above, we get:
 \begin{eqnarray}
 \label{24}
 &2& c(q^\mu) - 2 b_\mu(q^\mu) \partial^\mu S - i\hbar \partial^\mu b_\mu (q^\mu) \nonumber \\&+& a(q^\mu)( \partial_\mu S \partial^\mu S - i \hbar \partial_\mu \partial^\mu S )\nonumber\\&-&i\hbar \partial^\mu a(q^\mu) \partial_\mu S - \frac{\hbar^2}{4}\partial^\mu \partial_\mu a(q^\mu)=0
 \end{eqnarray}
  By separating real and imaginary parts and to introduce $ a(q^\mu)=\frac{\alpha(q^\mu)}{2m} $ we have:
 \begin{equation}
 \label{25}
 \frac{\partial _\mu S \partial^\mu S}{2m}+ \frac{2 c(q^\mu)}{\alpha ( q^\mu)}- \frac{\hbar^2}{8m}\frac{\partial^\mu \partial_\mu \alpha (q^\mu)}{\alpha (q ^\mu)}-\frac{2 b_\mu (q^\mu)\partial^ \mu S}{\alpha(q^\mu)}=0
 \end{equation}
 \begin{equation}\label{26}
 \partial ^\mu b_\mu ( q^\mu) + \partial ^\mu \left( \alpha(q^\mu)\frac{\partial_\mu S}{m}\right)=0.
 \end{equation}
 In comparison with RHJ and continues equations, which are obtained by setting polar form of wave function into Klein-Gordon (KG) equation \cite{nikolic}:
 \begin{eqnarray}
 \label{kg1}
 &&\frac{\partial^\mu S \partial_\mu S}{2 m_0}-\frac{m_0 c^2}{2}-\frac{\hbar^2}{2 m_0}\frac{\partial^\mu \partial_\mu R}{R}=0 \nonumber\\\\
 \label{kg2}&&\partial^\mu(R^2 \partial_\mu S)=0,
 \end{eqnarray}
 we conclude that  $ \alpha (q^\mu)=R^2(q^\mu)$, $ c(q^\mu)=-\frac{1}{4}m_0^2 c^2 R^2(q^\mu)$ and $ \partial^\mu b_\mu (q^\mu)=0$ where $  b_\mu (q^\mu)= \vec{V}(q^\mu) R^2(q^\mu)/2m$; as a consequence for $\vec{V}(q^\mu)=0$, so Eqs. (\ref{kg1}) and (\ref{kg2}) take the form:
 \begin{eqnarray}
 \label{kg3}
 &&\frac{\partial^\mu S \partial_\mu S}{2 m_0}-\frac{m_0 c^2}{2}-\frac{\hbar^2}{4 m_0}\frac{\partial^\mu \partial_\mu R}{R}-\frac{\hbar^2}{4m_0}\left(\frac{\partial_\mu R}{R}\right)^2=0\nonumber\\ \\
  &&\partial^\mu(R^2 \partial_\mu S)=0.
 \end{eqnarray}
 The results are same as non-relativistic ones. Again in this case, we see that half of the relativistic quantum potential and half of the relativistic quantum kinetic energy are appeared. We suppose that the spontaneous appearance of these terms leads to breaking their effects in half. In the other words, vacuum fluctuation's share in particles's energy has been split.
\section{Conclusion}
 Quantum potential is the term that has provoked some hostile criticism to the Bohm approach. Heisenberg  himself called it ad hoc, a sentiment that is  repeated in \cite{polking}. In this work, we came to the theory's defense by providing an innovative approach to QP via HJ equation, quantum commuting equations, quantum canonical transformations and without straight effect of wave function. The ultimate outcome includes QP and the new term, we called quantum kinetic energy (QK). What is more important for us is the spontaneous appearance of "quantum potential" (QP) and "quantum kinetic energy" ( QK) via a mere quantum approach, without the straight effect of wave function and its independence to the (polar) form of wave function. So we have proved that QP is not ad hoc and is not merely a reflection of the wave function \cite{maudlin} but along with   are produced due to the non-commuting operators in the HJ equation. Besides, along with other HJ equation terms, QP and QK are independent entities and are not dependent on the wave function form or wave equations formula.\\
More to the point, in both non relativistic and relativistic approach, half of QP and QK has been obtained. Since both QP and QK are due to vacuum fluctuations \cite{bohm, grossing8}, we suppose the coincident appearance of these terms leads to the breaking their effects in half. In  other words, vacuum fluctuation's share in particle's energy has been split. Also we suppose the automatic emergence of the quantum kinetic energy, may lead to another proof for "the vacuum fluctuation theorem" (VFT) \cite{grossing8}. According to VFT, additional kinetic energy could be a good candidate for "the vacuums zero-point energy"\cite{grossing8}. Furthermore we guess   this term may help   to describe causal interpretation of quantum relativistic phenomena.
\section{Acknowledgment}
Authors are deeply indebted to Prof. Mehdi Golshani for his encouraging interest in this work by useful discussions and answering our questions during the initial stages. We acknowledge the strong cooperation and useful discussions received by Dr. Majid Saberi Fathi. The authors thank Dr. Mehdi Atigh and Mr. Jalal Yousofi for the useful discussions and friendly support. Also M.M and M.R are grateful to the Institute for Research in Fundamental Sciences (IPM) for their kind hospitality. At last but not least authors thank Dr. Gerhard Grossing, Dr. Harvoj Nikolic for careful reading of the initial draft of this paper and their helpful points and Dr. L. S. Schulman for useful discussions.


\begin{thebibliography}{99}
\bibitem{holland} P. R. Holland, 1993, The Quantum Theory of Motion, Cambridge: Cambridge University Press.
\bibitem{bohm} D. J. Bohm, B. J. Hiley, 1982, foundation of Physics, $\mathbf{12}$, No 10, 1001-1016.
\bibitem{elitzur} A. Elitzur, S. Dolvo, N. Kolenda, 2005, Quo Vadis Quantum Mechanics?, Springer.
\bibitem{hiley} B. J. Hiley, 2002, Proc. Conf. Quantum Theory: reconsideration of foundations. V$\ddot{a}$xj$\ddot{o}$ University press.
\bibitem{polking} J. C. Polkinghorne, 2002, Quantum theory: a very short introduction, Oxford University press.
\bibitem{jordan} P. Jordan, Z. Phys. $\mathbf{37}$, 383 (1926); $\mathbf{38}$, 513 (1926).
\bibitem{goldstein} H. Goldstein, C. Poole, J. Safko, 2002, Classical Mechanics, $3^{rd}$ ed. Addison-Wesley.
\bibitem{roncadelli} M. Roncadelli  L.S. Schulman, Phys. Rev. Lett.$\mathbf{99}$,170406 (2007).
\bibitem{maudlin} T. Maudlin, 2002, Quantum Non-Locality and Relativity: Metaphysical Intimations of Modern Physics (Aristotelian Society Monographs), $2^{nd}$ed. Blackwell Publishing.
 \bibitem{provonic} S. Provonic,  eprint arXiv: 1005.4217
\bibitem{aharanov}Y. Aharonov and D. Bohm, ``Time in the Quantum Theory and the Uncertainty Relation for
Time and Energy," Phys. Rev. 122, 1649-1658 (1961).
\bibitem{pauli}W. Pauli, $\emph{Die allgemeinen Prinzipien der Wellenmechanik}$, edited by S. Fl$\ddot{u}$gge, Encyclopedia
of Physics (Springer-Verlag, Berlin, 1958), Vol. V, pt. 1, pp.1-168. See footnote 1 on p. 60.

\bibitem{wang} Z. Y. Wang and C. D. Xiong, Annals of Physics, Vol. 322, No 10, 2304-2314(2007).
\bibitem{miyamoto}M. Miyamoto, J. Math. Phys. $\mathbf{42}$, 1038 (2001).
\bibitem{razavi} M. Razavy, Amer. J. Phys. $\mathbf{35}$, 955 (1967).
\bibitem{nikolic} H. Nikoli$\acute{c} $, Found.Phys.Lett. $\mathbf{18}$, 549-561 (2005).
\bibitem{explain} The possible sets of independent canonical variables are: $ (\hat{q},\hat{Q}),(\hat{q},\hat{P}),(\hat{Q},\hat{p}),(\hat{p},\hat{P})$
\bibitem{lee} Lee, T. D., 1981, Particle Physics and Introduction to Field Theory, Harwood Acad. Pub., New York, 476-480.
\bibitem{grossing} G. Grossing, Physics A. $\mathbf{388}$, Issue 6, (2008) 811-823.
\bibitem{rovelli} C. Rovelli, 2004, Quantum Gravity, Cambridge University Press.
\bibitem{grossing8} G. Grossing, Phys. Lett. A, $\mathbf{327}$, issue 25 (2008) 4556-4563.

\end{thebibliography}
\end{document}